\documentclass[12pt]{article}
\usepackage{a4wide}
\usepackage{amsfonts}

\begin{document}

\begin{flushright}
{}FTUV 06/2712 \quad IFIC 06-48
\end{flushright}
\vskip0.5cm

\begin{center}
\begin{Large}
{\bf  BPS preons and higher spin theory in $D=4, 6,
10$\footnote{Invited contribution delivered at the Wroclaw  XXII
Max Born Symposium, September 27-29, 2006} }
\end{Large}

 \vskip1cm

{\bf Igor A. Bandos${}^{a,b}$ and Jos\'e A. de Azc\'arraga${}^a$}
 \vskip.5cm
 {\it ${}^a$ Dept.Theoretical Physics, Valencia University, \\and
IFIC (CSIC-UVEG), Spain\footnote{\small bandos@ific.uv.es,$\,$
j.a.de.azcarraga@ific.uv.es}}\\
{\it ${}^b$ ITP  KIPT Kharkov, Ukraine}

\end{center}

\begin{abstract}
We briefly review here the notion of BPS preons, the hypothetical
constituents of M-theory, emphasizing its generalization to
arbitrary dimensions $D$ and its relation to higher spin theories
in $D=4,6$ and 10.
\end{abstract}

\section{BPS preons in M-theory and supergravity ($D$=$11$)}

In $D$=11, {\it BPS preons} \cite{BPS01} {\it are M-theory BPS
states preserving all but one of the $32$ supersymmetries},
$|BPS\; preon>= |31/32 \; BPS>$. This implies that there exist
{\it 31} bosonic spinors $\epsilon_I^{\;\alpha}$,
 such that
\begin{eqnarray}\label{Preons}
 \fbox{$ \epsilon_I^{\;\alpha} Q_{\alpha}|BPS\; preon>= 0\;$} \;   ,
 \qquad \alpha= 1,\ldots , 32\; , \quad  I=1, \ldots , 31\quad ,
\end{eqnarray}
where the $Q_{\alpha}$'s are the $32$ supersymmetry generators of
M-theory. The $\epsilon_I^{\;\alpha}$ characterize the 31
preserved supersymmetries given by $\varepsilon=\kappa^I
\epsilon_I{}^\alpha$, where $\kappa^I$ are fermionic parameters,
and correspond to the Killing spinors in the supergravity
description.

Equivalently, a BPS preon may be characterized by {\it one}
bosonic spinor $\lambda_\alpha$ such that
\begin{eqnarray}\label{Pr-l}
  \fbox{$  Q_{\alpha}|BPS\; preon>\; \propto \; \lambda_\alpha\; $}\;   \;
  \Rightarrow  \; & |BPS\; preon>= |{\it 31\over 32} \; BPS>= |\lambda > \; .
\end{eqnarray}
The {\it preonic spinor} $\lambda_\alpha$ is clearly orthogonal to
the $31$ Killing spinors in (\ref{Preons}),
\begin{eqnarray}\label{epXl=0}
 \epsilon_I^{\;\alpha} \lambda_\alpha= 0\;  ,
 \qquad \alpha= 1,\ldots , 32\; , \quad  I=1, \ldots , 31
  . \qquad
\end{eqnarray}

The preonic nature of the $31/32$ states comes from the fact
\cite{BPS01} that a $k/32$-supersymmetric BPS state can be
considered as a composite of $\tilde{n}=32-k$ different BPS
preons, schematically
\begin{eqnarray}\label{composed}
& |{k\over 32} \;  BPS>   = | \lambda^{(1)}> \otimes \ldots
\otimes
|\lambda^{(32-k)}> \equiv \bigoplus\limits_{l=1}^{\tilde{n}=32-k}| \lambda^{(l)}>  \qquad  \\
& \qquad =|preon ^{(1)}> \otimes |preon ^{(2)}> \otimes \ldots
\otimes |preon ^{(32-k)}> \quad ,
 \nonumber
 \end{eqnarray}
and it is characterized by $\tilde{n}$ preonic spinors
$\lambda^{(l)}$ orthogonal to the $k$ Killing spinors. The
completely supersymmetric, $32/32$ BPS states, which are usually
identified with {\it supersymmetric vacua, do not contain any
preons}. Adding a preon to a $k/32$ state one obtains a state
breaking one more supersymmetry, $(32-k)+1$ in all. Thus {\it
preons can be thought of as the fundamental constituents of
M-theory} \cite{BPS01} (see \cite{IgorPLB03,30/32,BPS03,BPSrev}
for further discussion). Building a more composite BPS state from
a given one corresponds to breaking of one or more of the
originally preserved supersymmetries. In this picture, the fully
non-supersymmetric states appear as the most complicated ones:
they are composites of the maximal number, $32$, of independent
preons.

The $k$-supersymmetric M-theory BPS states, $|k/32\, BPS>$, are
usually associated with supersymmetric solutions of its low energy
limit, which is identified with the $D=11$ or $D=10$ type II
supergravities. The most important solutions were considered to be
the $1/2$ ones, corresponding to $|16/32 \; BPS>$ states, which
contain the $D=10$ Dirichlet $p$-branes and the $D=11$ M-branes
(see \cite{Duff94}). The less than 1/2 supersymmetric states were
identified with the intersecting branes (see \cite{Kallosh+=97}).
Thus, before the discovery (in 2002) of solutions preserving more
than 1/2 supersymmetries ($k/32 > 1/2$; see refs. in
\cite{Duff:02-06,BPS03,GGPR-r-06}), the main classification of the
M-theory BPS states was based on their (intersecting) brane
contents, and so it included the 1/2 BPS states and the
intersecting branes when $k/32 < 1/2$. The 2001 preonic conjecture
\cite{BPS01} provided an alternative algebraic classification of
all possible BPS states in terms of their preons contents. It
included all the then known BPS states and allowed for the
existence of any $k/32$ states (see also \cite{BL98} and
\cite{G+H99}). In this sense it predicted the appearance of the
supergravity solutions preserving more than 1/2 of the
supersymmetries.

Recent studies seem to indicate there is, at least at the
classical level, a preon `conspiracy' precluding the existence of
preonic solutions, both in IIB \cite{GGPR-IIB-06} and IIA
\cite{BdAV06} supergravities. According to \cite{GGPR-11-06}, the
simply-connected $31/32$, preonic solutions of $D$=11 supergravity
are forbidden as well. This preon conspiracy in classical
supergravity is nevertheless compatible with the BPS preon
conjecture since preons were rather introduced as M-theoretic
objects \cite{BPS01}. One might then establish a parallel between
this preon conspiracy and quark confinement, and state that only
composites of a certain number $\tilde{n}_{min}$ of BPS preons can
be `observed' as supergravity solutions. An interesting point for
future study is whether including quantum (`stringy' or
`M-theoretic') corrections in the supergravity equations would
change the situation and allow for the existence of preonic
solutions (see \cite{BdAV06} for further discussion).

Another question concerns the level of this preon conspiracy in
classical supergravity {\it i.e.}, the minimal (but nonzero)
allowed number $\tilde{n}=32-k$ of preons which can form the BPS
state described by a supergravity solution. By the above
discussion, such a solution would preserve $k=32-\tilde{n}$ out of
32 supersymmetries. Solutions preserving up to $k=28$
supersymmetries are known in IIB and up to 26 in IIA and $D$=11
supergravities (see refs. in \cite{Duff:02-06,BPS03,BdAV06}).
Thus, the problem is whether $\tilde{n}=2$ and $3$ solutions,
which preserve $30/32$ and $29/32$ supersymmetries respectively,
do exist. Recent work \cite{GGS06} excludes the existence of
solutions describing {\it two}-preon states in $D=5$ and in $D=4,
N=2$ supergravity. The existence of $\tilde{n}=2$ solutions in
$D$=10,11 is still open.

\section{ D=4,6 and 10 BPS preons and free conformal higher spin theories}

The notion of BPS preon applies \cite{BPS01,30/32} to an arbitrary
number of spacetime dimensions $D$. In $D=4,6$ and 10 a preon
state would be a BPS state preserving $3/4$, $7/8$ and $15/16$ of
the supersymmetries, respectively. In these `stringy' dimensions a
{\it BPS preon is related to the} $D=4,6$ and $10$ {\it free
massless conformal higher spin theory}.

To exhibit this relation explicitly, we first notice that an
equivalent (to (\ref{Pr-l})) definition of the BPS preon is given
by the following relation \cite{BPS01}
\begin{eqnarray}\label{Pr-Pll}
  \fbox{$  P_{\alpha\beta}|BPS\; preon>=
  \lambda_\alpha\; \lambda_\beta |BPS\; preon> $}\;  , \quad
  \alpha\, , \beta =1, \ldots , n\;\;   \; , \quad
\end{eqnarray}
which implies that the preon eigenvalues matrix of the generalized
momentum $P_{\alpha\beta}$ is of rank {\it one}. The
$P_{\alpha\beta}$ are the abelian bosonic generators of the
general supersymmetry algebra
 \begin{eqnarray}\label{M-alg} {} \{
Q_\alpha \, , \, Q_\beta \} = 2P_{\alpha\beta}\; , \qquad [
P_{\alpha\beta} \, , \, P_{\gamma \delta}] = 0\; , \qquad \alpha
  =1, \ldots , n\quad ,
 \end{eqnarray}
which for $n=32$ gives the M-algebra corresponding to $D$=11 (or
$D$=10 type II with the proper index interpretation). The
generators of this superalgebra have a natural differential
operator representation
\begin{eqnarray}
\label{P=d}
& P_{\alpha\beta}=  i {\partial \quad \over \partial
X^{\alpha\beta}}=: i
\partial_{\alpha\beta}\; , \qquad Q_{\alpha}= \, i \, {\partial\;\;  \over \partial
\theta^{\alpha}} + \, \theta^{\beta}\,
\partial_{\alpha\beta}\; , \qquad
 \end{eqnarray}
on an enlarged {\it tensorial superspace} parametrized by $n$
fermionic and $n(n+1)/2$ bosonic coordinates (see {\it e.g.}
\cite{30/32}),
\begin{eqnarray}\label{Sigma}
\Sigma^{({n(n+1)\over 2}|n)} \; : \quad \{ (X^{\alpha\beta} \, ,
\, \theta^\alpha) \} \; , \quad X^{\alpha\beta} \, = \,
X^{\beta\alpha} \, \; , \qquad \alpha \, , \beta=1, \ldots , n  \;
. \quad
 \end{eqnarray}
In addition to the $D$ vectorial, spacetime coordinates $x^a = (1/
n)\, X^{\alpha\beta} \, \Gamma^a_{\alpha\beta} \,$,
$\Sigma^{({n(n+1)\over 2}|n)}$  contains a number of tensorial
coordinates, $y^{[D/2]} \, \propto \, X^{\alpha\beta} \,
\Gamma^{a_1\ldots a_{[D/2]}}_{\alpha\beta}$, the types and number
of which depend on $n$ and $D$. For standard (one-time) spacetimes
(see \cite{Itzak+MP05} and refs. therein for two-time physics), $
X^{\alpha\beta }$ contains
\begin{eqnarray}\label{Xab=}
{} \{ X^{\alpha\beta } \} = \cases{\matrix{
  x^a \; , \quad & for \; n=2\, , \, D=3  \quad &  (\ref{Xab=}\; a) \cr
 (x^a, \, y^{ab}) \, \; , \quad & for \; n=4\, , \, D=4 \quad &  (\ref{Xab=}\; b) \cr
 (x^a, \, y_I^{abc}) \, \; , \; I=1,2,3 \quad  & for \; n=8\, , \, D=6 &  (\ref{Xab=}\; c)
 \cr
 (x^a, \, y^{abcde}) \, \; , \quad  & for \; n=16\, , \, D=10 &  (\ref{Xab=}\; d) \cr
 (x^a \, , \, y^{ab}\, , \, y^{abcde} \, ) \, \; ,\quad  & for  \;
      n=32\, , \, D=11 &  (\ref{Xab=}\; e)  }} {}
 \end{eqnarray}
For $n$=2, $X^{\alpha\beta}= X^{\beta\alpha}$ just provides
another presentation of the $D$=3 spacetime coordinates. For
$n\geq 4$ further bosonic coordinates appear, as {\it e.g.} 6 in
$D=4$ ($n$=4), $126$ in $D$=10 ($n$=16) and $517$=528-11 for
$D$=11 ($n$=32).

In $D$=4, the superspace $\Sigma^{(10|4)}$  of (\ref{Xab=}b) was
proposed in \cite{Fronsdal86} as a basis for description of
massless higher spin theories (see \cite{hirev}). A dynamical
realization of these ideas was found \cite{BLS99} to be given by a
generalized superparticle model, the $\Sigma^{({n(n+1)\over
2}|n)}$  tensorial superparticle \cite{BL98}. Its action  reads
\begin{eqnarray}\label{preonSSP}
S&:=& \int d\tau \, \lambda_\alpha(\tau)\lambda_\beta(\tau) \,
\Pi_\tau^{\alpha\beta} := \int d\tau \, \lambda_\alpha
\lambda_\beta (\partial_\tau X^{\alpha\beta} - i
\partial_\tau \theta^{(\alpha} \theta^{\beta )})
\; . \qquad
  \end{eqnarray}
This is generalization (to $n >2$, $\alpha , \beta =1, \ldots ,
n$) of the $D$=3 version of that of the Ferber-Schirafuji
superparticle action \cite{F78S83}. Indeed, the classical
mechanics counterpart of the second definition of BPS preon, Eq.
(\ref{Pr-Pll}),
\begin{eqnarray}\label{preonCM}
\fbox{$ \mathbb{P}_{\alpha\beta}
-\lambda_\alpha(\tau)\lambda_\beta(\tau) \approx 0 $} \, ,\qquad &
 \mathbb{P}_{\alpha\beta}:=  {\partial L\over \partial (\partial_\tau
 X^{\alpha\beta})}
\;  \qquad
  \end{eqnarray}
follows from (\ref{preonSSP}) as a primary constraint. The
$\Sigma^{({n(n+1)\over 2}|n)}$ superparticle model
(\ref{preonSSP}) possesses  $(n-1)$ $\kappa$--symmetries
\cite{kappa-s} and $n$ supersymmetries. This implies \cite{BL98}
that its ground state preserves all but one of the tangent space
supersymmetries and, thus, can be identified \cite{BPS03,30/32}
with a BPS preon.

 Upon a quantization (which converts second
class constraints into first class ones) \cite{BLS99} the
constraint (\ref{preonCM}) is imposed on the wave function in the
coordinate representation. Ignoring here fermionic coordinates for
simplicity, this gives the {\it preonic equation}
\cite{30/32,BPST05}
\begin{eqnarray}
\label{preonicEq}
(\partial_{\alpha\beta} + i \lambda_\alpha(\tau)\lambda_\beta(\tau))
\, G(X,\lambda) = 0 \quad , \quad
  \end{eqnarray}
which admits the `plain wave' solution $G(X,\lambda)=
\phi(\lambda)\, e^{- i \lambda_\alpha\lambda_\beta
X^{\alpha\beta}}$. Clearly, the bosonic spinor $\lambda_\alpha$
carries the (generalized) momentum degrees of freedom so that the
(generalized) coordinate representation for the wavefunction is
given by the integral of $G(X,\lambda)$ on $\lambda$ for some
measure of integration. The simplest one, $d^{n} \lambda$, gives a
scalar wavefunction $b(X)=\int d^n\lambda\, G(X,\lambda)= \int
d^n\lambda \, \phi(\lambda)\, e^{- i \lambda_\alpha\lambda_\beta
X^{\alpha\beta}}$; choosing alternatively $d^{n} \lambda \,
\lambda_\alpha$ one arrives at a spinor wavefunction suitable for
describing fermions, $f_\alpha (X)=\int d^n\lambda \,
\lambda_\alpha {\tilde G}(X,\lambda)= \int d^n\lambda \,
\lambda_\alpha \, {\tilde \phi}(\lambda)\, e^{- i
\lambda_\alpha\lambda_\beta X^{\alpha\beta}}$. These wavefunctions
obey the equations
\begin{eqnarray}\label{b}
\partial_{\alpha\beta}
\partial_{\gamma\delta}\,b(X)-\partial_{\alpha\gamma}\partial_{\beta\delta}\,b(X)&=&0\;,\\
\quad \partial_{\alpha\beta} f_\gamma(X)-\partial_{\alpha\gamma}
f_\beta(X)&=&0 \label{f} \; ,
\end{eqnarray}
which were proposed by Vasiliev \cite{V01s} to describe $D$=4
massless higher spin theories. These were generalized to
(enlarged) $AdS$ superspaces ($OSp(1|n)$ supermanifolds)
\cite{Misha+03}, and were shown to describe a whole tower of
bosonic and fermionic free massless {\it conformal} higher spin
fields also in $D$=6,10 \cite{BBdAST04}.

The field strength of the spacetime higher spin fields can be
extracted, {\it e.g.}, by decomposing the $b(X)=b(x,y)$ and
$f_\alpha(x,y)$ in a power series on the tensorial coordinates,
$y^{[D/2]}$ ($y^{[2]} =y^{mn}$, $y^{[3]} =y_I^{mnk}$ and $y^{[5]}
=y^{mnklp}$ for $D=4,6,10$, Eqs. (\ref{Xab=}b)--(\ref{Xab=}d)).
Schematically (see \cite{BBdAST04, BPST05} for the precise
expressions),
\begin{eqnarray}
\label{ymnpqr} b(x,\,y)=&\phi(x)+y^{ [D/2]}F_{ [D/2]}(x) +y^{
[D/2]_1}\,y^{[D/2]_2}\, R_{[D/2]_1 [D/2]_2}(x)+ \qquad \nonumber
\\ & \qquad + \sum_{s=3}^{\infty}\,y^{ [D/2]_1}\cdots
y^{[D/2]_s}\,{R}_{[D/2]_1\, \,\cdots\, \,[D/2]_s}(x)\,, \nonumber \\
 f_\alpha(x,\,y)
 =&\psi_\alpha(x)+y^{ [D/2]} {{\mathbf{\Psi}}}_{\alpha\,[D/2]}(x) + y^{ [D/2]}y^{ [D/2]'}
 {{\mathbf{\Psi}}}_{\alpha\,[D/2]\, [D/2]'}(x)+... \qquad \, .
\end{eqnarray}

As is well known, in $D$=4 all the free massless equations are
conformally invariant. Consequently, all the massless field
strengths are included in the $n=4$ version of the decomposition
(\ref{ymnpqr}) on $y^{ [2]}=y^{mn}$. In particular, $F_{[2]}:=
F_{mn}(x)=-F_{nm}(x)$ is the field strength of Maxwell field,
${R}_{m_1n_1\, m_2n_2}(x)={R}_{m_2n_2\, m_1n_1}(x)$ is the
linearized Riemann tensor, ${\mathbf{\Psi}}^\alpha_{[2]}(x
):={\mathbf{\Psi}}^\alpha_{[mn]}(x )$ is the Rarita-Schwinger
field strength (spin $3/2$), etc. Eqs. (\ref{b}) and (\ref{f}) fix
both the algebraic properties of these and other higher spin field
strengths  (such as the Bianchi identities $R_{[m_1 n_1\; m_2]
n_2}=0$ for the linearized Riemann tensor) and also define the
linear differential equations for these field strengths (such as
the Bianchi identities for the Maxwell field strength and the
equations of motion) \cite{V01s,Misha+03,BBdAST04}.

In contrast, in $D$=6 and 10 not all the massless fields are
conformal and, consequently, not all massless fields but only the
conformal ones enter the decomposition (\ref{ymnpqr}) for $n$=8
and $16$. In $D$=10 these are, in addition to the usual scalar and
spinor fields $\phi(x)$ and $\psi_\alpha(x)$, the basic self-dual
five form field strength $F_{[m_1m_2\ldots m_5]}= {1\over 5!}
\epsilon_{m_1m_2\ldots m_5n_1n_2\ldots n_5}F^{[n_1n_2\ldots n_5]}$
(characteristic of type IIB supergravity) and the tensors with
several symmetrized groups of `five's' {\it i.e.}, with
symmetrized sets of five antisymmetric self-dual indices,
$R_{[5]_1 [5]_2}= R_{[5]_2 [5]_1}$ etc., as well as their
fermionic counterparts \cite{BBdAST04}.

 The bosonic $b(X)$ and fermionic $f_\alpha(X)$ fields
are the two lowest components of a superfield on the
$\Sigma^{({n(n+1)\over 2}|n)}$ superspace, $\Phi(X\, , \theta)=
b(X) + \theta^\alpha f_\alpha(X) + {\cal O}(\theta \, \theta )$.
Then the free conformal higher spin equations (\ref{b}) and
(\ref{f}) follow from the simple linear differential equation
\cite{BPST05}
\begin{eqnarray}\label{DDF=0} D_{[\alpha} D_{\beta ]} \Phi
(X\; , \theta)=0\; , \quad & D_{\alpha}:= \,  {\partial\;\;  \over
\partial \theta^{\alpha}} + {i}\, \theta^{\beta}
\partial_{\alpha\beta } \qquad \; .
\end{eqnarray}
A calculation shows that Eq. (\ref{DDF=0}) also implies the
vanishing of all higher components of $\Phi(X\; , \theta)$. The
group-theoretical meaning of this equation was discussed in
\cite{ZhI+JL05}, while its curved space (generalized AdS)
generalization and supergravity in tensorial superspace was the
subject of \cite{BPST05} to which we also refer for a discussion
on the problems and perspectives for an interacting higher spin
theory in this framework.

In the same way as the scalar bosonic and the spinor fermionic
wavefunctions, $b(X)$ and $f_\alpha(X)$ in Eqs. (\ref{b}),
(\ref{f}), are constructed from the solution of the preonic
equation (\ref{preonicEq}) \cite{30/32}, one can express the
solution of the superfield equation (\ref{DDF=0}) as an integral
$\Phi (X\; , \theta) =\int d^n\lambda {\cal G}_0 (X, \theta,
\lambda)$ of the $\lambda$--dependent (`phase space') superfield
${\cal G}_0 (X, \theta, \lambda)$ which obeys the following {\it
superfield generalization} \cite{BPST05} of the preonic equation
(\ref{preonicEq})
\begin{eqnarray}\label{preonsusy}
& (D_{\alpha}D_{\beta }-  2\lambda_\alpha\lambda_\beta)\, {\cal
G}_0 (X, \theta,\lambda ) = 0 \; .
\end{eqnarray}
The antisymmetric part of this equation gives rise to
(\ref{DDF=0}) while the symmetric part has the form of the preonic
equation (\ref{preonicEq}) ($\{ D_\alpha\, , \, D_\beta\} = 2i
\partial_{\alpha\beta}$). The phase space superfield ${\cal G}_0
$, in its turn, appears as the leading component of the {\it
Clifford superfield} $\mathbb{G} (X, \theta,\lambda \, , \chi )=
{\cal G}_0 (X, \theta,\lambda ) + \chi {\cal G}_1 (X, \theta,\lambda
)$, $\chi \chi =1$  \cite{Dima88}, which obeys the first order
Clifford superspace equation \cite{BLS99,BPST05}
\begin{eqnarray}
\label{preonPre-susy} & (D_{\alpha}+{i}\chi\,  \lambda_\alpha)\,
{\mathbb G} (X, \theta,\lambda, \chi ) = 0 \; , \qquad \chi\chi =1
\; .
\end{eqnarray}

 Eq. (\ref{preonPre-susy}) implies $D_{\alpha} {\cal G}_0 +
i\lambda_\alpha {\cal G}_1 =0$ and $D_{\alpha} {\cal G}_1 +
i\lambda_\alpha {\cal G}_0 =0$. These mean, besides that both
components ${\cal G}_0$, ${\cal G}_1$ of the Clifford superfield
${\mathbb G}$ obey the preonic equation (\ref{preonicEq})
(representing in tensorial spacetime the second definition
(\ref{Pr-Pll}) of a BPS preon), that $Q_{\alpha} {\cal G}_0
\propto \lambda_\alpha$ and $ Q_{\alpha} {\cal G}_1 \propto
\lambda_\alpha$ are valid. {\it Any} of these two equations
provide a tensorial superspace representation (see Eq.(\ref{P=d}))
of the first definition (\ref{Pr-l}) of a BPS preon.

\section{Concluding remarks: fermionic preons?}

The above discussion suggests the possibility of considering BPS
preons with not only bosonic, but also with fermionic (and,
perhaps, even with exotic) statistics; this requires further
study. Here we only notice that the two equivalent definitions of
a BPS preon, Eqs.(\ref{Pr-l}) and (\ref{Pr-Pll}), imply the
existence of an ultrashort {\it preonic supermultiplet} containing
one bosonic and one fermionic state, $|\lambda \;, b>$ and
$|\lambda \;, f>$, characterized by the {\it same} bosonic spinor
$\lambda_\alpha$, such that
\begin{eqnarray}\label{Preon1+1-susy}
& Q_\alpha |\lambda \;, b > = \lambda_\alpha |\lambda  \;, f >\;
,\qquad Q_\alpha |\lambda\;,  f > = \lambda_\alpha |\lambda \;, b
>\; .  \qquad
\end{eqnarray}
For their associated fields $\phi (\mathbb{X}) = <
\mathbb{X}|\lambda \,, b>$ and $\psi (\mathbb{X}) =
<\mathbb{X}|\lambda \,, f >$ (where $\mathbb{X}$ may be
$(X^{\alpha\beta},\,\lambda_\alpha)$ or different), the
supersymmetry transformations in (\ref{Preon1+1-susy}) read
$\,\delta \phi(\mathbb{X})= \varepsilon^\alpha \lambda_\alpha
\psi(\mathbb{X})$,  $\delta \psi(\mathbb{X})= \varepsilon^\alpha
\lambda_\alpha \phi(\mathbb{X})$. As a ground state is taken to be
bosonic, $\psi(\mathbb{X})=0$, such a state is clearly invariant
under the $31$ supersymmetries associated with the $31$ Killing
spinors $\epsilon_I{}^\alpha$ of Eq. (\ref{epXl=0}). This bosonic
ground state configuration is identified with a BPS preon.
However, one sees that the same $31$ supersymmetries are preserved
by the purely fermionic ($\phi(\mathbb{X})=0$) state characterized
by the Grassmann odd function $\psi(\mathbb{X})$, the fermionic
counterpart of the bosonic BPS preon. It would be interesting to
understand whether such a simple algebraic construction of a
fermionic BPS preon also has a dynamical realization.

{\bf Acknowledgments}. The authors thank Jos\'e Izquierdo, Jurek
Lukiersky and Dima Sorokin for their collaboration on the subject
and for numerous useful discussions. Partial support from research
grants from the Spanish MEC and EU FEDER funds (FIS2005-02761),
the Generalitat Valenciana, INTAS-2005-1000008-7928, the EU RTN
Network MRTN-CT-2004-005104 and the Ukrainian State Fund for
Fundamental Research (383) is also acknowledged.

{

}

\end{document}